\documentclass[12pt,preprint]{aastex}

\begin{document}

\slugcomment{Accepted for publication in ApJ Letters}

\title{Optical Interferometric Observations of $\theta^{1}$ Orionis C
  from NPOI and Implications for the System Orbit}

\author{J. Patience\altaffilmark{1}, R.~T. Zavala\altaffilmark{2}, 
L. Prato\altaffilmark{3}, O. Franz\altaffilmark{3}, L. Wasserman\altaffilmark{3}, 
C. Tycner\altaffilmark{4}, D.~J. Hutter\altaffilmark{2}, C.~A. Hummel\altaffilmark{5}}

\altaffiltext{1}{University of Exeter, School of Physics, Astrophysics Group,
Stocker Road, Exeter, EX4 4QL United Kingdom; patience@astro.ex.ac.uk}  
\altaffiltext{2}{U.S. Naval Observatory, Flagstaff Station, 10391 W. Naval
Obs. Rd., Flagstaff, AZ 86001; bzavala@nofs.navy.mil; djh@nofs.navy.mil}
\altaffiltext{3}{Lowell Observatory, 1400 West Mars Hill Rd.,
  Flagstaff, AZ 86001; lprato@lowell.edu; otto.franz@lowell.edu; lhw@lowell.edu}
\altaffiltext{4}{Department of Physics, Central Michigan University, Mt. Pleasant, 
MI 48859; c.tycner@cmich.edu}
\altaffiltext{5}{European Southern Observatory, Karl-Schwarzschild-Str. 2, 
85748 Garching, Germany; chummel@eso.org}

\begin{abstract}
With the Navy Prototype Optical Interferometer (NPOI), the
binary system $\theta^{1}$ Orionis C, the most massive member of the
Trapezium, was spatially resolved over a time period extending from
February 2006 to March 2007. The
data show significant orbital motion over the 14 months, and, after
combining the 
NPOI data with previous measurements of the system from the
literature, the observations 
span 10 years of the orbit. Our results indicate that the
secondary did not experience an unusually close periastron passage this year,
in contradiction to the prediction of a recently published, highly
eccentric $\sim$11 year orbit. Future observations 
of this source will be required to improve the orbital solution. Possible
implications of the results in terms of system distance
are discussed, although a main conclusion of this work 
is that a definitive orbit solution will require more time to obtain 
sufficient phase coverage, and that the interaction 
effects expected at periastron did not occur in 2007.   
  
\end{abstract}

\keywords{binaries: close; open clusters and associations: individual
  (Trapezium); stars: individual ($\theta^{1}$ Orionis C); techniques:
  interferometric}

\section{Introduction}

Orion is the nearest example of a giant molecular cloud and is the
site of both high-mass star formation and a prodigious number of
recently-formed stars; the central 2.5 pc region of the Orion Nebula
Cluster (ONC) alone contains $\sim$3500 stars with a combined mass exceeding
900 $M_{\odot}$ \citep{hil97}. Although star-forming regions such as
Taurus are closer, these regions are dark clouds associated with
low-mass star formation and far fewer total stars. Initial star-count
studies \citep{lad91} implied that up to 95\% of stars formed in
clustered environments like Orion, while more recent {\it Spitzer
Space Telescope} results indicate that at
least half of all stars originate in dense regions \citep{meg04}.
An analysis of binary star
distributions \citep{pat02} suggests that approximately 70\% of field
stars may have formed in a clustered environment.
Thus, the study of Orion, the closest giant star forming cluster, is
central to investigating the early history of the majority of stars.
Furthermore, the variety of cluster morphologies investigated with
{\it Spitzer} observations suggest
that OB stars play a significant role in the formation and evolution
of star formation in clusters \citep{meg04}.

The distance to Orion is a critical parameter that influences the
interpretation of many of the properties of the region and its
members. Measurements range from $\sim$390 pc to $\sim$480 pc
using a variety of methods and assumptions including observations of water
masers, radio sources, an eclipsing binary, and statistical analysis 
\citep[and references therein]{gen81, sta04, san07, men07, jef07}.
A closer distance would imply that the
stars are less luminous, and members that are still contracting onto the
main sequence are consequently older based on comparison with
theoretical evolutionary tracks \citep[e.g.,][]{sie00, pal99}.
Older ages for the stars above the main sequence
suggest a spread in ages.

Binary stars yield
a model-independent distance with the combination of a
spatially resolved orbit and a double-lined radial velocity orbit
\citep[e.g.,][]{tor97}. Given the distance to Orion, very high
angular resolution is required to separate binary systems which exhibit
significant orbital motion. The binary
system $\theta^{1}$ Ori C (HD 37022) in the
Trapezium region of Orion has such a close separation that it was not
detected until speckle observations barely resolved the pair
\citep{wei99} with a separation of less than the diffraction limit of
the 6m telescope employed in the discovery.
The separation has decreased over time and is now best 
monitored with interferometry; a recent orbit fit
suggested the system might be just past periastron and completing an
orbital cycle within a year \citep{kra07}. In this letter, we present
the results of interferometric observations with NPOI that add
significantly to the binary orbit phase coverage and suggest the orbital
period may be substantially longer than predicted.

The distance to $\theta^{1}$ Ori C in particular and its physical
parameters such as mass and age are of great importance
since the O7 primary \citep[e.g.,][]{gar82},
as the most massive member of the ONC, has the
dominant role in shaping the properties of the surrounding nebula, and
strongly impacts the circumstellar material around the ONC stars. The
photoionizing radiation from $\theta^{1}$ Ori C produces the brightening
of many proplyds \citep{ode93}, but also causes the material to
escape. Observations of mass loss rates \citep{joh98, hen99} and
theoretical modeling of the results \citep[e.g.,][]{sto99} imply 
that these structures cannot survive for $\ga10^5$ yrs, substantially 
less than the $<$1$-$2 Myr age of the ONC \citep{hil97}. Possible explanations
for the apparent contradiction in disk lifetimes and stellar ages
include radial orbits for the proplyds \citep{sto99} or a very young
age for $\theta^{1}$ Ori C, such as has been proposed for $\theta^{1}$ 
Ori B \citep{pal01}. In contrast, recent models of the evolution of
disk sizes in the ONC \citep{cla07} suggest that disk survival times of 1-2 Myr in
the uv field produced by $\theta^{1}$ Ori C are possible -- consistent
with the age of the stellar population. Key to
estimating the age of $\theta^{1}$ Ori C is placing the secondary
accurately on the H-R diagram with a well-measured distance,
luminosity, and mass given that the primary has already contracted
onto the main sequence. We present new magnitude differences
which will aid in the assessment of the luminosity, but concentrate on the
orbital motion which is required to estimate the distance.

\section{Observations and Data Reduction}

During Feb. 2006 to Mar. 2007, $\theta^{1}$ Ori C was observed with NPOI
on 6 nights. The NPOI array at
Anderson Mesa near Flagstaff, AZ,  
consists of six 50 cm (12 cm clear aperture) siderostats which can be
deployed along a Y-shaped array of
vacuum light pipes \citep{arm98}. The wavelength coverage spans
550$-$850 nm over 16 spectral channels. Details regarding the NPOI observational
setup and data recording can be found
in \citet{hum03} and \citet{ben03}. 
More recent upgrades relevant to this program are improvements in the 
acquisition camera sensitivity and calibration of the bias level for
low count rates; implementation of these two
changes enabled the observations of $\theta^{1}$ Ori C. Table 1 provides 
a log of the NPOI observations.

The observations of $\theta^{1}$ Ori C were interleaved with
$\epsilon$ Ori, one of the bright (V=1.70) belt stars. The visibilities
from $\epsilon$ Ori serve to calibrate both variable atmospheric
conditions and the system response to a point source. The small angular
diameter of $\epsilon$ Ori -- 0.86 $\pm$ 0.16 milli-arcseconds (mas)
\citep{moz91}
-- and its proximity to  
$\theta^{1}$ Ori C -- less than 5\arcdeg\ separation -- allowed for
accurate atmospheric 
correction. For both the target and calibrator, the fringes were
recorded in 2 ms frames for a total scan duration of 
30 s before switching to the other source. The
individual 2 ms data were averaged over a 1 s time period, and these
1 s data points were checked for outliers before averaging to
generate 30 s averaged $V^{2}$ measurements. Calibration factors
were determined by comparing the observed data 
of $\epsilon$ Ori to that expected from a 0.86 mas diameter single star. 
These calibration factors were then applied to observations of
$\theta^{1}$ Ori C. 
The flagging, averaging, and
calibration steps were performed with the OYSTER package, as described
in Hummel et al. (1998, 2003), except that the bias corrections were 
determined for each star individually. 

\section{Results and Analysis}

The calibrated visibilities were fit with a model comprised of two
stars with slightly resolved stellar surfaces. The primary star diameter 
was estimated to be 0.3 mas using the expected 
diameter of an O7 star at a distance of 450 pc \citep{dri00}. Assuming the spectral type from
\citet{kra07}, the secondary 
diameter was set to half that of the primary, rounded to 0.2 mas. The
observational setup did not 
allow us to fit for such small diameters convincingly, so we held these 
values constant. Figure 1 shows examples
of calibrated $V^2$ values and the best fit model. Predicted visibilities
from a recent orbit solution \citep{kra07} are also plotted. Table 1
lists the epoch, siderostats 
used, number of baselines, maximum baseline length, number of scans,
estimated separation and position angle with uncertainties, and the
position error ellipses. 
Because each scan yields up to 16 $V^2$ values per baseline and up to 405
$V^2$ points were obtained during a night, individual 
measurements are not listed, but examples are plotted in Figure
1. Model fits to individual baselines include a 180$\arcdeg$ ambiguity in the
position angle, and the values listed in Table 1 are chosen based on
previous measurements and the inability to fit orbits if the companion
was located in the opposite quadrant in the 2007 data.
Some nights listed in Table 1 included closure phase observations, but we 
defer discussion of these results, which have the ability to directly
resolve the 180$\arcdeg$ ambiguity, for a future paper including results from
calibration binaries.

While the bandpass does
not exactly match the $V$ or $R$ filters, our magnitude 
difference measured at 550 nm closely approximates that of the V band \citep{zav07}. 
Considering all the data, the best estimates for the observed magnitude
differences are $\Delta$ mag$_{\rm 550 nm} = 1.3 \pm 0.3$ and
$\Delta$ mag$_{\rm 700 nm}= 1.2 \pm
0.2$. The NPOI measurement agrees with previous $\Delta V$
estimates from speckle observations taken at the 6.0~m Special
Astrophysical Observatory at Mt. Pastukhov in Russia \citep[1.1 mag,][]{kra07}.

Orbits based on the previous measurements and the new NPOI data
are shown in Figure 2. The earliest NPOI data from Feb. 2006 show
only minimal 
orbital motion from the IOTA Dec. 2005 data \citep{kra07}, indicating there is not
an offset between the two systems. The 2007 NPOI data show significant
evolution in the orbit. The measured separations of the secondary
relative to the primary are larger than expected from the predicted
orbit and lag behind the solution in orbital phase \citep{kra07}. The
NPOI results indicate that a very close
periastron passage did not take place this year as
suggested by the preliminary orbital solutions of \citet{kra07}.

As indicated in Figure 2, only part of the orbit -- probably less than
one half -- is covered by the extant data, making any assessment of
the orbit fit preliminary. Combining all previous position
measurements and associated error bars with the NPOI visibilities and
uncertainties, we fit an orbit using
ORBGRID \citep{hmf89, mdh99} and used its solution
as a  starting point for a least squares
solution; uncertainties from the covariance matrix of
the least squares solution are quoted (Table 2).  Both
ORBGRID and the least squares solution
weight the astrometric points in a relative sense,
and we set these weights according to the areas of the 
error ellipses from our Table 1 and Table 3 of \citet{kra07}.

We stress that the orbital elements of the best fit based on the
current data (Table 2) may be modified
substantially as more data become available. Compared with the
previous solution, we find a longer period and a much less eccentric
orbit. Our lower eccentricity of
0.16 is well within the bulk of the distribution for T Tauri binaries
\citep[e.g.,][]{mat94}, the 
lower mass counterparts of $\theta^1$ Ori C. In contrast, the earlier
solution found an extremely high 
eccentricity of 0.91$-$0.93 (Kraus et al. 2007).

Although it is premature to calculate a robust
distance to the Orion Nebula Cluster from our data, if we
assume a total system mass of 40~M$_{\odot}$ the orbital elements
in Table~\ref{orbit} give a dynamical parallax distance of 730 pc ---
unrealistically large considering the
distance to the background high extinction molecular gas
\citep{gen81}. The 40~M$_{\odot}$ value is the minimum mass estimate
from the evolutionary tracks of \citet{wal94}, using a T$_{\rm eff}$ for an
O7 star of 36,000~K \citep{mas05}.  The values of $\Delta$mag$_{\rm 550nm}$ and
$\Delta$mag$_{\rm 700nm}$ between the primary and secondary stars imply at the
latest a B2 secondary spectral type \citep{dri00}, with a corresponding
T$_{\rm eff}$ of 28,000~K.  Thus, 40~M$_{\odot}$ is a lower limit; larger
masses would imply an even greater distance.

Figure 3 summarizes our dynamical parallax
measurement.  The uncertainty in the value of 
a$^3$/P$^2$ (Table~\ref{orbit}), directly related to parallax, does not
yield a reliable estimate of distance at this time.
Increasing the semi-major axis
by 1 $\sigma$ and reducing the period by 1 $\sigma$ yields a low value
for the distance of 344 pc for the same total mass.  Clearly, the
available data do not significantly constrain the distance to $\theta^1$ Ori C.
Figure 3 compares the best-fit semi-major axis and period, with their
associated 1 $\sigma$ and 2 $\sigma$ error ellipses, to
the values expected for the two Orion distances estimated from the
radio star and the maser.  To explore fits to our data with more physical
distances imposed, we selected a range of nine periods from 10 to 26 years
and determined the corresponding semi-major axes for distances of 390~pc
and 480~pc. For both distances, fits were obtained for periods of $\sim$22
years with orbital elements that agreed to within $\sim3\sigma$ of the
best fit elements (Table~\ref{orbit}).  Fits for other periods were
significantly (by many sigma) worse.

Further monitoring of 
the orbit of $\theta^1$ Ori C is required to decrease the 
errors in the orbital elements and provide a reliable dynamical 
distance. Given the proximity to periastron passage,
continued observations are
particularly important over the next months and years.
Resolving the distance and mass of $\theta^1$ Ori C, and revealing
the nature of its interactions with the local environment,
will provide important insight into the closest region of
massive star formation.

\acknowledgments

The Navy Prototype Optical
Interferometer is a joint project of the 
Naval Research Laboratory and the US Naval Observatory, in
cooperation with Lowell Observatory, and is funded by the Office of
Naval Research and the Oceanographer of the Navy. The authors would
like to thank Jim Benson and the NPOI observational support staff Dave Allen,
Jim Clark, Brit O'Neill, Susan Strosahl, Dale Theiling, Josh Walton and 
Ron Winner whose efforts made this project possible. We thank Phil Massey, 
Nathan Mayne for helpful discussions, and Nat White for assistance 
with observing arrangements. RTZ thanks the Michelson Science Center 
for an invitation to visit which helped initiate this work, and JP
gratefully acknowledges funding from the Michelson Fellowship Program.

{\it Facilities:} \facility{NPOI ()}

\clearpage
 
\begin{deluxetable}{lccccccccccccc}
\rotate
\tabletypesize{\scriptsize}
\tablecaption{NPOI Observations and $V^{2}$ Model Fit Results \label{tbl-1}}
\tablewidth{0pt}
\tablehead{
\colhead{UT Date} & \colhead{Julian Year} & \colhead{Siderostats}   & \colhead{$\#$ b.l.}   &
\colhead{Max b.l.} & \colhead{$\# V^{2}$}  & \colhead{$\rho$}
& \colhead{$\sigma_{\rho}$} & \colhead{$\theta$} & \colhead{$\sigma_{\theta}$} & 
\colhead{$\sigma_{maj}$} & \colhead{$\sigma_{min}$} & \colhead{$\sigma_{\phi}$ (\arcdeg)} \\
\colhead{(1)} & \colhead{(2)} & \colhead{(3)} & \colhead{(4)} & \colhead{(5)} & 
\colhead{(6)} & \colhead{(7)} & \colhead{(8)} & \colhead{(9)} & \colhead{(10)} & 
\colhead{(11)} & \colhead{(12)} & \colhead{(13)} 
} 
\startdata
2006 Feb 24 & 2006.1486 & AC-AE-AW & 2 & 22.2 & 171 & 11.80 & 1.11  & 152.3 & 3.5 & 1.20 & 0.54 & 178.0 \\
2007 Feb 22 & 2007.1425 & AE-AW-W7 & 2 & 37.4 & 118 & 11.94 & 0.31 & 88.1 & 5.2   & 1.09 & 0.28 & 170.4 \\
2007 Feb 25 & 2007.1507 & AE-AW-W7 & 2 & 37.4 &  60 & 12.13 & 1.58 & 92.9 & 8.8   & 2.41 & 0.39 & 142.8 \\
2007 Mar 06 & 2007.1753 & AE-AW-AN-W7 & 4 & 38.1 & 405 & 12.17 & 0.37 & 86.6 & 2.1 & 0.46 & 0.36 & 157.6\\
2007 Mar 17 & 2007.2055 & AE-AW-AN-W7 & 4 & 38.1 & 135 & 12.28 & 0.41 & 85.6 & 1.9 & 0.46 & 0.35 & 42.2 \\
2007 Mar 20 & 2007.2137 & AE-AW-AN-W7 & 4 & 38.1 & 200 & 12.14 & 0.43 & 83.0 & 2.3 & 0.49 & 0.42 & 158.2\\
 
\enddata

 
\tablecomments{Col. (1): UT date of NPOI observation 
Col. (2): Julian Year of NPOI observation. Col. (3): Siderostats used Col. (4): Number of independent 
baselines Col. (5): Max. baseline length (m) Col. (6): Number of V$^2$ measurements Col. (7): Fitted 
binary separation (mas)  Col. (8): Error in $\rho$ (mas)  Col. (9): Fitted binary position angle 
(\arcdeg).; this angle is chosen as it is a smooth extension of
previous results and no orbital solutions could be found using $\theta+180\arcdeg$ Col. (10): Error in $\theta$ (\arcdeg) Col. (11): Semi-major axis of error ellipse (mas). 
Col. (12): Semi-minor axis of error ellipse (mas) Col. 13: Position angle of error ellipse.}
\end{deluxetable}

\clearpage

\begin{deluxetable}{cc}
\tablecaption{O{\sc rbital} E{\sc lements} \label{orbit}}
\tablewidth{0pt}
\tablehead{ \colhead{Data} & \colhead{Value}}
\startdata
a (mas)  & 41 $\pm$ 14 \\
i (deg)  & 107.2 $\pm$ 3.5 \\
$\Omega$ (deg) & 208.8 $\pm$ 3.7 \\
e        & 0.16 $\pm$ 0.14 \\
$\omega$ (deg)& 96.9 $\pm$ 118.7 \\
T$_0$ (JY) & 2007.0 $\pm$ 5.9 \\
T$_0$ (JD)   & 2454101 $\pm$ 2154 \\
P (days) & 9497 $\pm$ 1461 \\
P (years) & 26 $\pm$ 13 \\
a$^3$/P$^2$ (mas$^3$/yr$^2$) & 103 $\pm$ 146 \\
$\chi^2/dof$ & 2.6$\times10^{-6}$ \\
\enddata
\end{deluxetable}

\clearpage

\begin{figure}
\plottwo{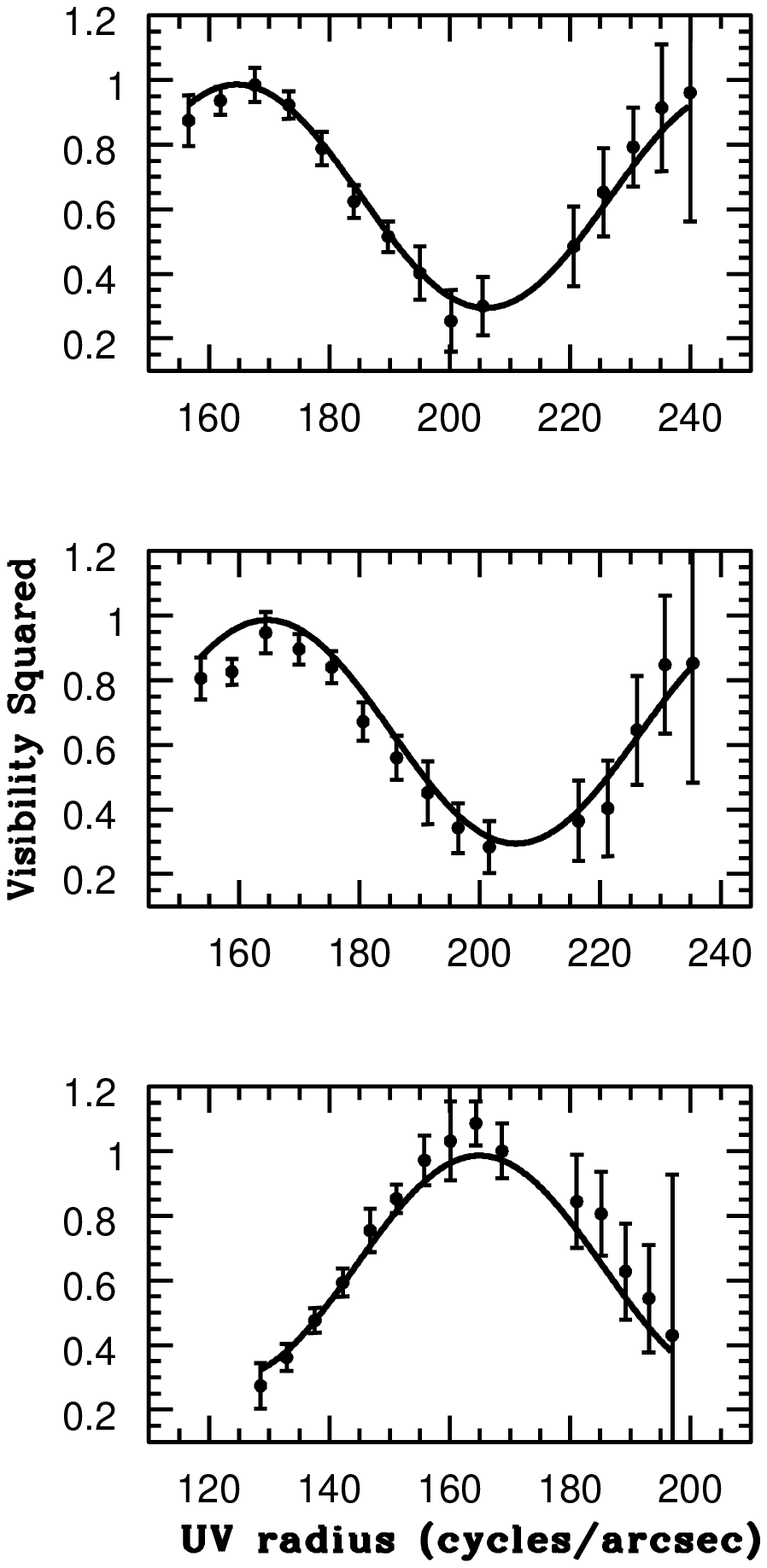}{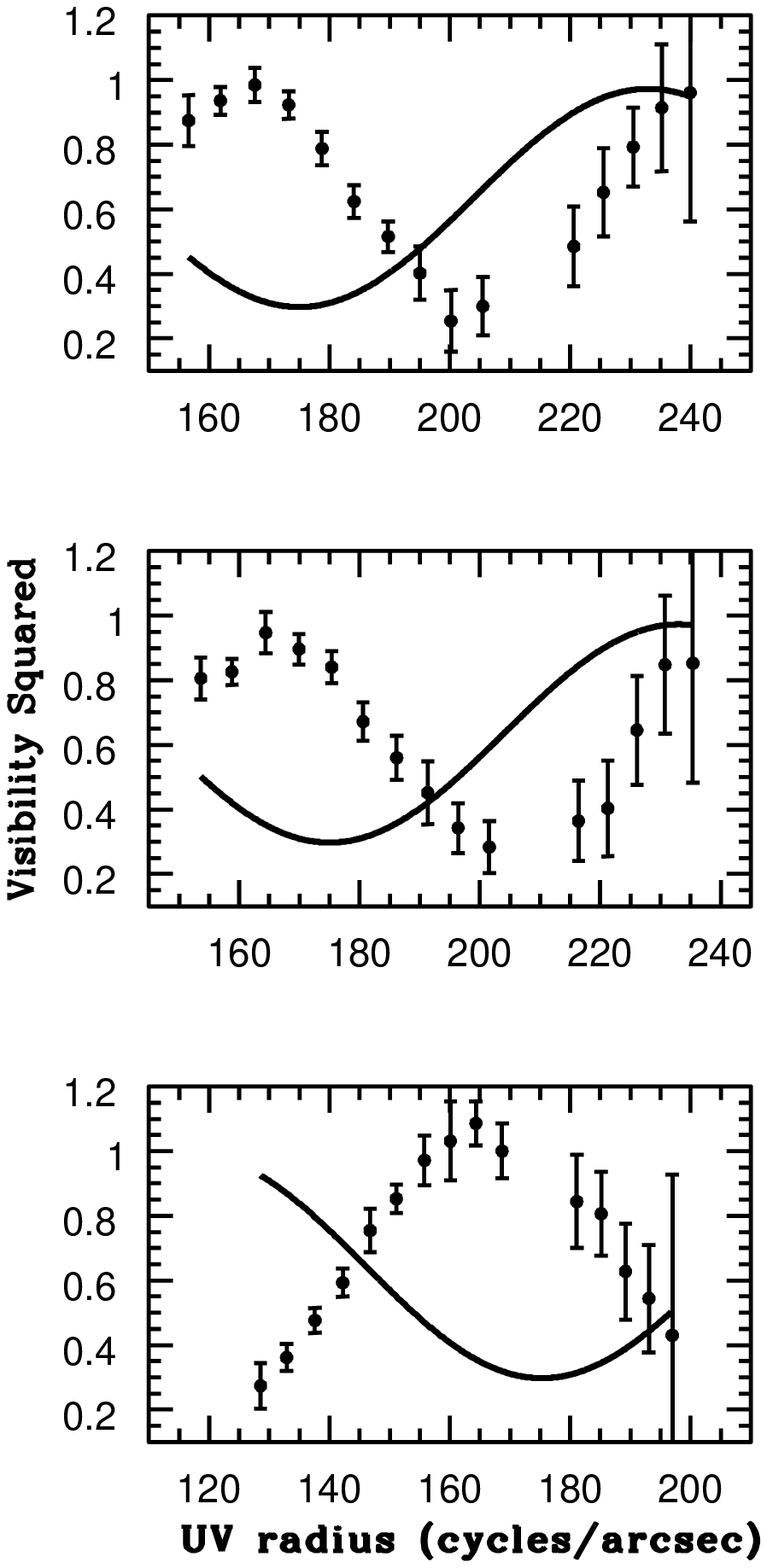}
\caption{Left: Panels showing 3 of 9 scans of the calibrated V$^2$ of 
$\theta^1$ Ori C observed with the AE-AW baseline on 2007 Mar 06 
overlaid with a model of separation  12.17~mas and position angle
86.6\arcdeg\ (Table 1). Right: A set of three panels shows the same
data but overlaid using the predicted separation of 8.6~mas at a
position angle of 86.28\arcdeg\ for orbit 1 of \citet{kra07} 
on that date. The errors include both a statistical error and an 
estimate of the uncertainty of the calibration. \label{f1}}
\end{figure}

\clearpage

\begin{figure}
\plotone{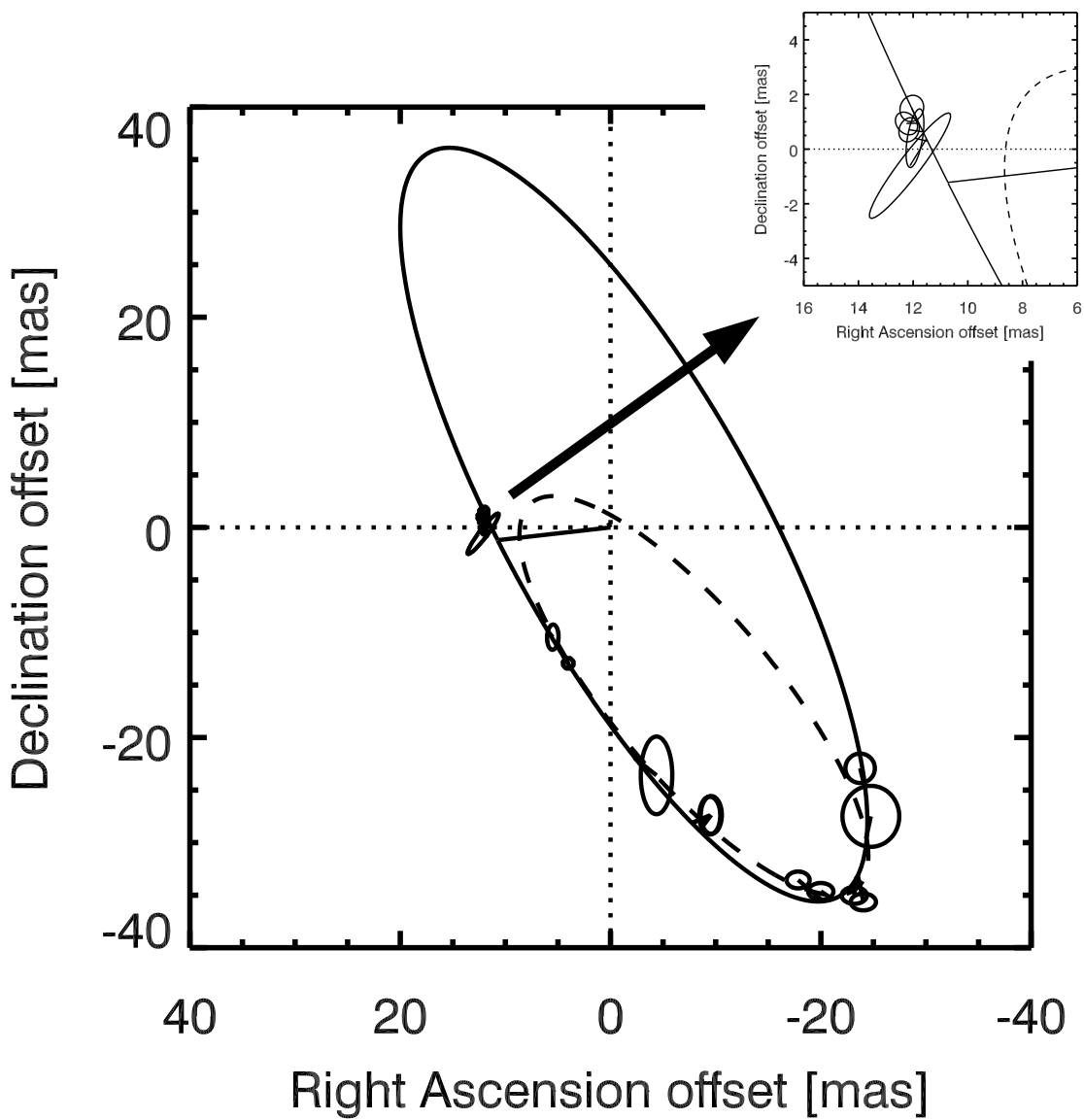}
\caption{The best-fit orbit of $\theta^1$ Ori C with orbital elements
  listed in Table~\ref{orbit} is plotted as a solid line. This fit
  incorporates the NPOI astrometric results from Table 1 and previous
  measurements given in \citet{kra07}. Error ellipses 
for the astrometric points are shown along with a vector indicating the 
periastron point. The dashed line shows the predicted orbit (Orbit1)
  from \citet{kra07}.  We display the closely spaced 2007 NPOI astrometric solutions 
in the inset for clarity. The next set of observations should
  demonstrate decisively the validity of a longer period solution. \label{f2}}
\end{figure}

\clearpage

\begin{figure}
\plotone{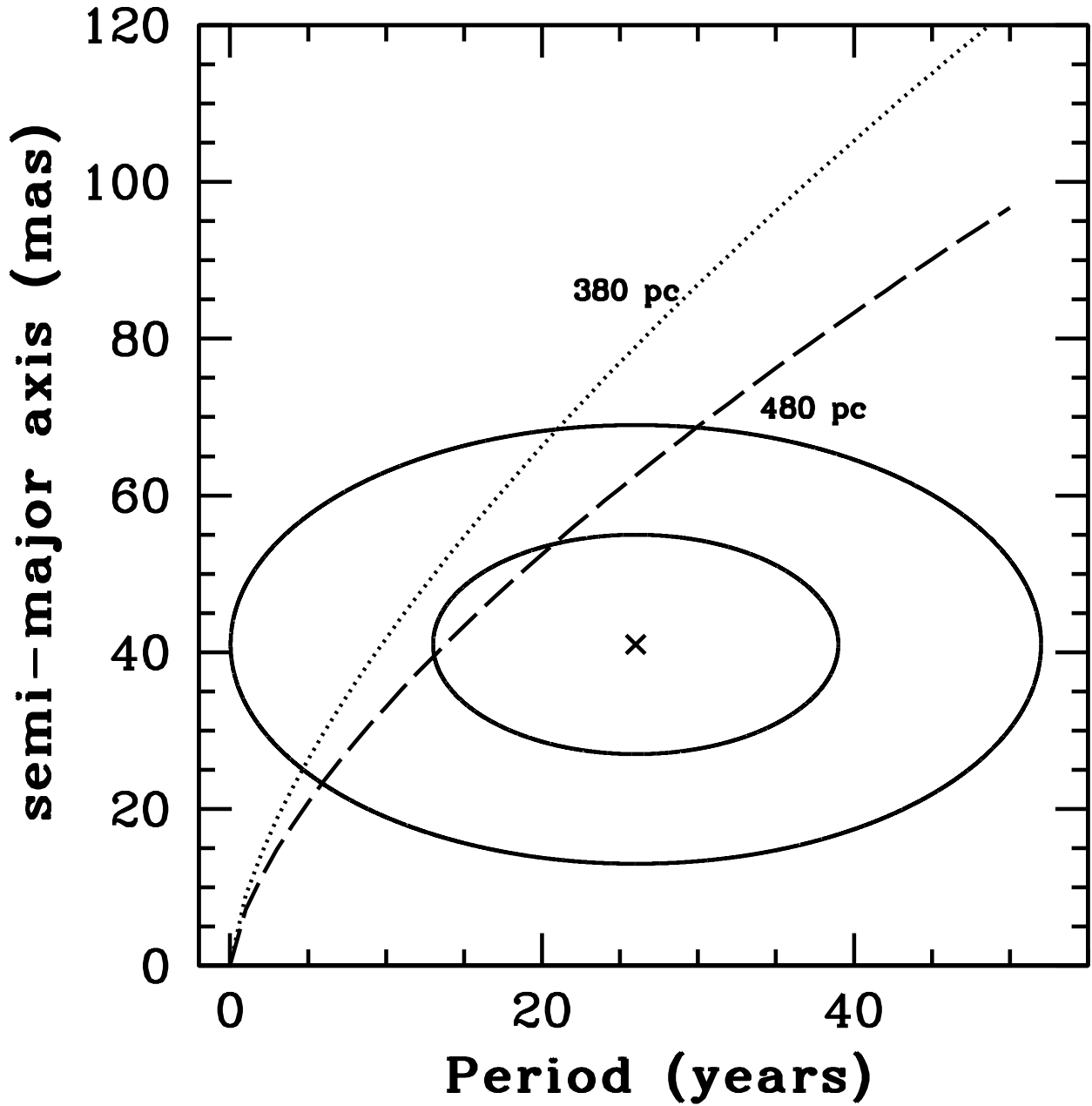}
\caption{The best-fit period and semi-major axis from Table 2 are
  plotted with a cross, and the superimposed error ellipses (solid
  lines) represent 
the 1 and 2 $\sigma$ errors of the orbital elements
in Table 2. For a total system mass of 40 M$_{\odot}$, the period and
  semi-major axis values consistent with distances of 480 pc (dashed)
  and 380 pc (dotted) are also indicated. At the $\sim$1 $\sigma$ level, the
  current orbit fit is consistent with essentially all measured
distances to $\theta^1$ Ori C. Given the limited
  phase  coverage of the orbit a definitive
  distance measurement will require continued observations. \label{f3}}
\end{figure}

\end{document}